\documentclass[prl,superscriptaddress,showpacs,twocolumn]{revtex4}%
\usepackage{graphicx}
\usepackage{amsmath}
\usepackage{amsfonts}
\usepackage{color}
\usepackage{amssymb}%
\usepackage{bm}
\usepackage{lipsum}
\usepackage{array}
\usepackage{longtable}
\begin{document}
\author{Guoren Zhang}
\affiliation{
Institute for Advanced Simulation, Forschungszentrum J\"ulich, 
D-52425 J\"ulich, Germany}
\author{Evgeny Gorelov}
\affiliation{
Institute for Advanced Simulation, Forschungszentrum J\"ulich, 
D-52425 J\"ulich, Germany}
\author{Esmaeel Sarvestani}
\affiliation{
Institute for Advanced Simulation, Forschungszentrum J\"ulich, 
D-52425 J\"ulich, Germany}
\author{Eva Pavarini}
\affiliation{
Institute for Advanced Simulation, Forschungszentrum J\"ulich, 
D-52425 J\"ulich, Germany}
\affiliation{JARA High-Performance Computing, RWTH Aachen University, 52062 Aachen, Germany}
\title{The Fermi surface of  Sr$_{2}$RuO$_{4}$: 
 spin-orbit and anisotropic Coulomb interaction effects}
\begin{abstract}
The topology of the Fermi surface of Sr$_{2}$RuO$_{4}$ is  well described
by local-density approximation calculations with spin-orbit interaction, but  the relative
size of its different sheets is not.
By accounting for many-body effects via dynamical mean-field theory, we show that the standard isotropic Coulomb interaction alone worsens or does not correct this discrepancy.
In order to reproduce experiments, it is essential
to account for the Coulomb anisotropy. The latter is small but has strong effects; it  competes
with the Coulomb-enhanced spin-orbit coupling and the isotropic Coulomb term
in determining the Fermi surface shape.
Its effects are likely sizable in other correlated multi-orbital systems. 
In addition, we find that the low-energy self-energy matrix -- responsible for the 
reshaping of the Fermi surface -- sizably differ from the static Hartree-Fock limit.
Finally,  we find
a strong spin-orbital {entanglement}; this supports the view  that the conventional description  of
 Cooper pairs via factorized spin and orbital part might not apply to Sr$_{2}$RuO$_{4}$.
\end{abstract}
\pacs{71.18.+y, 71.27.+a, 72.80.Ga, 71.10.-w}
\maketitle
%%%%%%%%%%%%%%%%%%%%%%%%%%%%%%%%%%%%%%%%%%%%%%%%%%%%%%%%%%%%%%%%%%%%%%
Sr$_2$RuO$_4$ has attracted a lot of attention as a possible realization of a spin-triplet  superconductor \cite{Rice95,Ishida,Mackenzie04,Maeno2012} and, at the same time, as a very peculiar strongly correlated metal 
\cite{Bergemann,Ingle,Schmidt,Nakatsuji98,Gorelov,hundness,malvestuto2011,malvestuto2013,stricker}.
Understanding the details of its Fermi surface\,(FS) is key to unravel the nature of quasi-electrons in the normal phase and  can cast light on the mechanism and the symmetry of the superconducting order parameter.
It is thus not surprising that the  Fermi surface of Sr$_2$RuO$_4$ has been 
intensively investigated, both experimentally \cite{Mackenzie01, Mackenzie02, Mackenzie03,Damascelli, D.H.Lu, Iwasawa, S.Y.Liu}
and theoretically \cite{Liebsch,Pavarini,Haverkort}.  
Although the main features are nowadays well understood,
the effects of the interplay between correlations, spin-orbit and crystal structure have not been  fully disentangled yet. 
%%%%%%%%%%%%%%%%%%%%%%%%%%%%%
\begin{figure}[t]
\begin{center}
\includegraphics[scale=0.42]{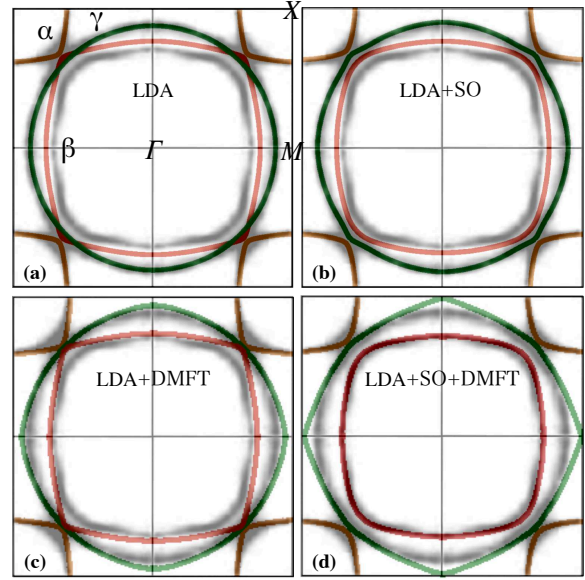}
\caption{(Color-online){\label{fig1} Fermi surface ($k_z\!=\!0$) of Sr$_{2}$RuO$_{4}$  from (a) LDA, (b) LDA+SO, (c) LDA+DMFT and {(d) LDA+SO+DMFT calculations performed with $O(3)$-symmetric Coulomb matrix}, $(U,J)=(3.1,0.7)$~eV, $T\to0$ limit. Light lines: $\alpha$ and $\beta$  sheets. Dark lines: $\gamma$  sheet. Grey density maps: experimental data taken from Ref.\,\onlinecite{Damascelli}.} }
\label{default}
\end{center}
\end{figure}  
%%%%%%%%%%%%%%%%%%%%%

Sr$_2$RuO$_4$ is a tetragonal layered perovskite  (space group I4/mmm \cite{Vogt}) with the Ru $4d^4$ ($t_{2g}^4 e_g^0$) electronic configuration and Ru atoms at sites with $D_{4h}$ symmetry;
due to the layered structure the  Ru $t_{2g}$ $xz$ and $yz$ bands are almost one-dimensional and very narrow,
with a band width $W_{xz}=W_{yz}$  about half as large as that of the two-dimensional Ru $xy$ band, $W_{xy}$. 
Experimentally, the Fermi  surface of Sr$_{2}$RuO$_{4}$ has been studied via  both the de Haas-van Alphen technique \cite{Mackenzie01,Mackenzie02,Mackenzie03} and angle-resolved photoemission spectroscopy\,(ARPES)\,\cite{Damascelli, D.H.Lu, Iwasawa, S.Y.Liu}.  
It is made (Fig.~\ref{fig1}) by  three sheets, the electron-like $\gamma$ ($xy$ band)
and $\beta$ ($xz,yz$ bands) sheets and the
hole-like $\alpha$ sheet  ($xz,yz$ bands). 
Theoretically, {\em ab-initio} calculations based on the local-density approximation (LDA) 
qualitatively reproduce the FS topology, provided that the spin-orbit (SO) interaction is taken into account \cite{Pavarini,Haverkort}. 
Indeed, 
several experiments point to a sizable SO coupling \cite{Rice95,Mackenzie2011,Veenstra}.
These calculations fail, however, in describing the relative size of the sheets,
suggesting that perhaps many-body effects play a key role.
The relevance of the Coulomb interaction for the electronic properties of 
Sr$_2$RuO$_4$, as well as its interplay with bands of different width,
was shown early on via model many-body studies \cite{Liebsch}.
More recently, LDA+DMFT (local-density approximation + dynamical mean-field theory)  calculations
have emphasized the interplay of Coulomb interaction and $t_{2g}$ crystal field (CF)  \cite{Liebsch2007,Gorelov}, and  the role of the Hund's rule coupling  \cite{hundness}.
LDA+slave-boson calculations point to SO effects on the correlated bands \cite{lechermann}.
It remains however unclear to what extent many-body effects actually modify the Fermi surface, and how they compete with other effects.

In this Letter, by using the LDA+DMFT method with SO interaction, we investigate, for the first time, the interplay between Coulomb repulsion,
spin-orbit and symmetry at the Fermi surface in a realistic setting. We show that, surprisingly, the standard isotropic Coulomb interaction alone ($O(3)$ symmetry) does not improve (or even worsens) the agreement between theoretical and experimental Fermi surface. 
The agreement with experiments can only be achieved if both SO and  Ru $D_{4h}$ low-symmetry Coulomb terms are taken into account. These terms are often neglected
in realistic many-body calculations due to the numerical difficulties of treating them.
In order to efficiently deal with 
many-body Hamiltonians  of arbitrary symmetry we have recently developed a generalized  LDA+DMFT  solver  \cite{Gorelov,flesch,Zhang} based on the continuous-time quantum Monte Carlo  \cite{ctqmc} technique. Here  we use the interaction-expansion \cite{Rubtsov}
 flavor (CT-INT) of this solver \cite{Gorelov}, further extended to account for SO terms. 
We show that, remarkably, 
$D_{4h}$ low-symmetry Coulomb terms compete 
with the standard isotropic $O(3)$  terms,  the crystal-field and the SO coupling in determining the actual 
shape of the FS of Sr$_2$RuO$_4$.

In the first step we perform LDA calculations  using the full-potential linearized augmented plane-wave  
method (Wien2k\,\cite{Blaha} code), with and without SO interaction. Next we construct localized $t_{2g}$ Wannier functions
via Marzari-Vanderbilt localization  \cite{MarzariVanderbilt,Mostofi} and $t_{2g}$ projectors \cite{notewannier}.
Finally we build the $t_{2g}$ Hubbard model 
%%%%%%%%%%%%%
\begin{eqnarray}\label{ham}
H&=&-\sum _{j j'} \sum_{\sigma \sigma'}\sum_{mm'}t_{m\sigma,m'\sigma'}^{j,j'}c_{jm \sigma}^{\dagger}c_{j' m'  \sigma'}\\ 
&+&\nonumber \sum_j\sum_{\sigma\sigma'}\!\!\sum_{ mm' p p'} \!\! \!\!U_{mm'p p'} c^\dagger_{jm\sigma} c^\dagger_{jm'\sigma^\prime} c_{jp'\sigma^\prime} c_{jp\sigma} -H_{\rm dc}
\end{eqnarray}
%%%%%%%%%%%%%
Here $c_{jm\sigma}^{\dagger}$ ($c_{jm\sigma}$)  creates (destroys) an electron with spin $\sigma$ in the Wannier state with orbital quantum number  $m$ ($m={xy},{yz},{xz}$) at site $j$; ${H}_{\rm dc}$ is the double-counting correction \cite{DC}; $-t_{m\sigma,m'\sigma'}^{j,j'}$ are the hopping integrals ($j\neq j'$) and the elements of the on-site energy matrix ($j=j'$). The latter includes crystal field splittings and, when present, the  SO term
$ {\bf l}\cdot \underline{\lambda}\cdot {\bf s}$, where $\underline{\lambda}$ is the coupling constant tensor. After ordering the states as  
$\{|m\rangle_{\uparrow}\}, \{|m\rangle_{\downarrow}\}$, the on-site matrix $\varepsilon_{m\sigma,m'\sigma'}=-t_{m\sigma,m'\sigma'}^{j,j}$ takes the form
 %%%%%%%%%%
\begin{eqnarray*}
 \hat{\varepsilon}=  \left( \begin{array}{cccccc} 
                                         \varepsilon_{xy} &0&0&0&\frac{\lambda_{xy}}{2}&\frac{i\lambda_{xy}}{2} \\[.4ex] 
                                         0& \varepsilon_{yz}&-\frac{i\lambda_{z}}{2} &-\frac{\lambda_{xy}}{2}&0&0\\[.4ex]
                                         0&\frac{i\lambda_{z}}{2} & \varepsilon_{xz}&-\frac{i\lambda_{xy}}{2} &0&0\\[.4ex]
                                         0&-\frac{\lambda_{xy}}{2}&\frac{i\lambda_{xy}}{2} & \varepsilon_{xy}&0&0\\[.4ex]
                                         \frac{\lambda_{xy}}{2}&0&0&0& \varepsilon_{yz}&\frac{i\lambda_{z}}{2} \\[.4ex]
                                         -\frac{i\lambda_{xy}}{2} &0&0&0&-\frac{i\lambda_{z}}{2} & \varepsilon_{xz}
                                 \end{array} \right).                                                                                                                                                                                                                                            
 \end{eqnarray*} 
 %%%%%%%%%%%%%%%%%%%%%%%%%%%%%%%%%%%%
Due to $D_{4h}$ site symmetry, the Ru $t_{2g}$ states split into an $e_g$ doublet $(xz,yz)$ and a 
$b_{2g}$ singlet $(xy)$, with on-site energy $\varepsilon_{xz}=\varepsilon_{yz}$  and $\varepsilon_{xy}$, respectively.
LDA yields $\varepsilon_{xz}=\varepsilon_{xy}+\varepsilon_{\rm CF}$ with $\varepsilon_{\rm CF}\sim 120$~meV.
The SO parameter $\lambda_{z}$ couples the orbital $|yz\rangle_{\sigma}$ to the orbital $|{xz}\rangle_{\sigma}$; instead, the term $\lambda_{xy}$ couples the $|{xy}\rangle_{\sigma}$ state to the $|{yz}\rangle_{-\sigma}$ and
$|xz\rangle_{-\sigma}$ orbitals. 
LDA yields  $\lambda_{z}\sim 102$~meV and  $\lambda_{xy}\sim 100$\,meV, i.e., 15\% smaller than the value
$130\pm30$\,meV estimated via spin-resolved photoemission spectroscopy \cite{Veenstra}.
The LDA tetragonal anisotropy  $\delta_\lambda=\lambda_{z}-\lambda_{xy}$, is tiny, $\delta_\lambda \sim 2$ meV.
The terms $U_{mm'pp'}$ are elements of the screened Coulomb interaction tensor.
For a free atom the Coulomb interaction tensor for $d$ states can be written in terms of the three  Slater integrals $F_0$, $F_2$ and $F_4$. 
For $t_{2g}$ states the essential terms \cite{evacorrel11} are the direct 
($U_{mm'mm'} =U_{m,m'}=U-2J(1-\delta_{m,m'})$) and
 the exchange ($U_{mm'm'm}=J$) screened Coulomb interaction,  
the  pair-hopping ($U_{mmm'm'}=J$) and the spin-flip term ($U_{mm'm'm}=J$);
 here $U=F_0+\frac{4}{49}(F_2+F_4)$ and $J=\frac{1}{49}(3F_2+\frac{20}{9}F_4)$.
For site symmetry $D_{4h}$ the number of independent Coulomb parameters increases to six.
Here we will discuss in particular the effect of 
$\Delta U= U_{xy,xy}-U_{xz,xz}$ and $\Delta U^\prime=U_{xy,yz}-U_{xz,yz}$.
We solve  (\ref{ham}) with DMFT using CT-INT quantum Monte Carlo. We work with  a $6\times6$ self-energy matrix $\Sigma_{m\sigma,m'\sigma'}(\omega)=\Sigma^{'}_{m\sigma,m'\sigma'}(\omega)+i \,\Sigma^{''}_{m\sigma,m'\sigma'}(\omega)$ in spin-orbital space, extending the solver of Ref.~\cite{Gorelov} to deal explicitly with the SO term; 
$\Sigma^{'}$ is the real and $\Sigma^{''}$
the imaginary part of the self-energy. 
The calculations with SO coupling are performed in the basis $|\tilde{m}\rangle_\sigma=\hat{T} |m\rangle_\sigma$, where the unitary operator $\hat{T}$ is chosen such that the local imaginary-time Green function matrix is real. In the rest of the paper, for calculations with SO coupling, the elements of the self-energy matrix are given in the $|\tilde{m}\rangle_\sigma$  basis; since
$\hat{T}$ only changes the phases \cite{notephase} but does not mix orbitals, we rename for simplicity  $|\tilde{m}\rangle_\sigma$ as  $|{m}\rangle_\sigma$.

First let us analyze the LDA results without SO interaction (Fig.~\ref{fig1}(a)).
Our results agree very well with previous theoretical works\,\cite{Oguchi, Singh01, Pavarini, Haverkort}.  Compared with ARPES data, LDA describes well the $\alpha$ and $\gamma$ sheets, and in particular the region around the $M$ point of  the $\gamma$ sheet. 
 There are two major discrepancies. First, the LDA $\beta$ and $\gamma$ sheets cross,
 differently than in ARPES. Second, the area enclosed by the $\beta$ sheet is 
 larger in LDA than in ARPES. 
Once the SO interaction is switched on three relevant changes occur, as Fig.~1(b) shows. The $\beta$-$\gamma$ crossing  becomes an anti-crossing due to the SO coupling $\lambda_{xy}$; the $\beta$ sheet shrinks and  the $\gamma$ sheet expands.
These effects improve the overall agreement  \cite{Pavarini,Haverkort,Veenstra} with ARPES results, however, the $\beta$ sheet remains too  large with respect to experiments. 

The next step consists in incorporating the Coulomb interaction via LDA+DMFT  (see Fig.~\ref{fig1}(c)). First, we perform standard calculations with no SO term and  $O(3)$-symmetric Coulomb tensor. 
We use two sets of parameters: $(U,J)=(3.1,0.7)$~eV, as obtained via constrained LDA\,\cite{Pchelkina}, and $(U,J)=(2.3,0.4)$~eV, as obtained via
constrained random-phase approximation \cite{Mravlje} (cRPA).
{These sets of values yield spectral functions with Hubbard bands in line with available experiments \cite{Schmidt, Yokoya, Pchelkina, Kurmaev}.  %
Low-energy many-body effects change the splitting $\varepsilon_{\rm CF}$ into 
$\varepsilon_{\rm CF}+\Delta \varepsilon_{\rm CF}$, with \cite{zerofreq} 
  $$\Delta \varepsilon_{\rm CF} =  
\frac{1}{2}\Sigma^{\prime}_{yz\,\sigma,yz\,\sigma}(0)+\frac{1}{2}\Sigma^{\prime}_{xz\,\sigma,xz\,\sigma}(0)-\Sigma^{\prime}_{xy\,\sigma,xy\,\sigma}(0).$$  
The shift $\Delta \varepsilon_{\rm CF}$ turns out to be positive \cite{Liebsch};  at $T=290$\,K we find $\Delta\varepsilon_{\rm CF}={108}$\,meV 
for $(U,J)=(3.1,0.7)$~eV
 and ${80}$\,meV for $(U,J)=(2.3,0.4)$~eV. As a consequence, with respect to  LDA,  the $\alpha$ and $\gamma$ sheets expand  
  whereas the $\beta$ sheet shrinks. This is shown in Fig.~\ref{fig1}(c) for
$(U,J)=(3.1,0.7)$~eV; the LDA+DMFT Fermi surface deviates from ARPES in particular around the $M$ point ($\gamma$ sheet), which approaches the boundary of the first Brillouin zone. 
{For $(U,J)=(2.3,0.4)$~eV the effect is smaller and the FS remains closer to the LDA one \cite{suple}.}
%%%%%%%%%%%%%%%%%%%%%%%%%%%%%%%%%%%%
\begin{figure}[t]
\begin{center}
\includegraphics[angle=270,width=0.38\textwidth]{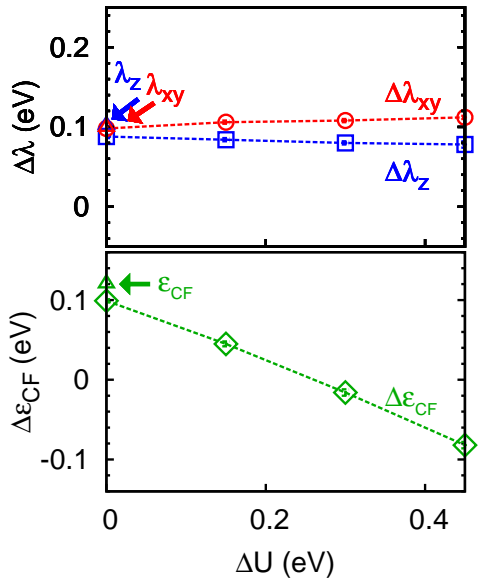}
\caption{\label{fig2} (Color-online) {Many-body corrections of the on-site parameters at the Fermi energy
as a function of $\Delta U$, with $\Delta U^\prime=\Delta U/3$. The LDA\,+\,SO\,+\,DMFT calculations are done at $T=290$\,K and for $(U,J)=(3.1,0.7)$~eV.
{Top:  Spin-orbit couplings corrections,
$\Delta \lambda_{z}$  and $\Delta \lambda_{xy}$.}
Bottom: Crystal-field splitting correction, $\Delta \varepsilon_{\rm CF}$.
The LDA values $\lambda_z$, $\lambda_{xy}$ and  $ \varepsilon_{\rm CF}$
are indicated by arrows.
QMC error bars are shown.} }
\label{default}
\end{center}
\end{figure}
%%%%%%%%%%%%%%%%%%%%%%%%%%%%%%%%%%%%

In Fig.~\ref{fig1}(d) we show the effect of including the SO term  (LDA+SO+DMFT).  
We find a $\Delta\varepsilon_{\rm CF}$ slightly smaller than for $\lambda_i=0$;  
the SO couplings are, however, sizably enhanced with respect to LDA,  i.e.,
$\lambda_{i}\to \lambda_{i}+
\Delta \lambda_{i}$, with
\begin{eqnarray*}
\Delta \lambda_{z}&=&-\bigg[\Sigma^{\prime}_{{yz}\uparrow, {xz} \uparrow}(0)+\Sigma^{\prime}_{{yz}\downarrow, {xz} \downarrow}(0)\bigg], \nonumber \\ 
\Delta \lambda_{xy}&=&\frac{1}{2}\sum_{\sigma}\sigma  \;\left[\Sigma^{\prime}_{{xy}\,\sigma, {yz} \,-\sigma}(0)-\Sigma^\prime_{{xy}\,\sigma, {xz} \,-\sigma}(0)\right].  
\end{eqnarray*}
{At 290~K we obtain $\Delta \lambda_{xy}\sim 96$~meV and $\Delta \lambda_{z}\sim 88$~meV}.
For the FS, with respect to LDA+DMFT, the agreement worsens for the $\gamma$ sheet and
it improves for the $\alpha$ and $\beta$ sheets.  The change can be ascribed to the enhanced SO couplings.
Comparing Fig.~\ref{fig1}(b) and (c) with Fig.~\ref{fig1}(d) 
it appears that the combined effect of Coulomb and SO interaction  results in the $\gamma$ sheet approaching the boundary of the first Brillouin zone. For $(U,J)=(2.3,0.4)$~eV
the effects of the SO coupling are qualitatively similar \cite{suple}.
These results point to the existence of an important mechanism  neglected so far. 

%%%%%%%%%%%%%%%%%%%%%%%%%%%%%%%%%%%%%%%%%%%%%%%%%%%%%%%%
\begin{figure}[t]
\begin{center}
\includegraphics[angle=270,width=0.38\textwidth]{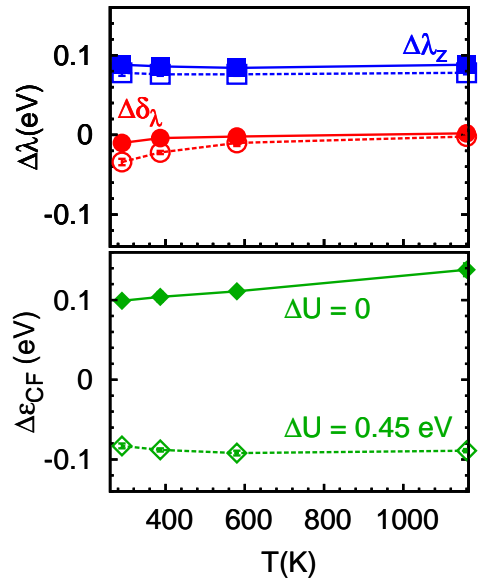}
\caption{\label{fig3}(Color-online) { 
Many-body corrections of the on-site parameters at the Fermi energy as a function of  temperature
and for $(U,J)=(3.1,0.7)$~eV.
{Top: Spin-orbit coupling corrections, $\Delta \lambda_{z}$ and 
$\Delta \delta_\lambda=\Delta \lambda_{z}-\Delta \lambda_{xy}$. }
Bottom: Crystal-field splitting correction $\Delta \varepsilon_{\rm CF}$.
Solid lines: $\Delta U=0$. Dashed lines: $\Delta U=0.45 $~eV. 
All calculations are performed for $\Delta U^\prime=\Delta U/3$. QMC error bars are shown.}}
\label{default}
\end{center}
\end{figure}
%%%%%%%%%%%%%%%%%%%%%%%%%%%%%%%%%%%%%%%%%%%%%%%%%%%%%%%%
We identify the missing mechanism in low-symmetry Coulomb terms. 
{Due to the elongation of the RuO bond in the {\bf c} direction,} the
$e_g$ (${xz,yz}$)  Wannier orbitals have a larger spread than ${xy}$ orbital \cite{Spread}, suggesting positive $\Delta U$ and $\Delta U^\prime$. This is in line with the results of cRPA, \mbox{$\Delta U\sim 0.3$\,eV\,\cite{Mravlje}}. 
To study the effect of the Coulomb anisotropy we perform two additional sets of  LDA+SO+DMFT  calculations,
the first with $0<\Delta U < 0.6$~eV and $\Delta U^\prime=0$
and the second with $0<\Delta U^\prime=\Delta U/3 < 0.2$~eV \cite{lastnote}. 
The most significant results  are shown in Fig.~\ref{fig2} for $T=290$\,K \cite{note}.{
{We find that both $\Delta \lambda_z$  and $\Delta \lambda_{xy}$ are weakly dependent on $\Delta U$.} 
}
Instead, $\Delta \varepsilon_{\rm CF}$ decreases linearly with $\Delta U$ and changes sign at a quite small $\Delta U\sim$\,0.25\,eV; at this value the effective CF has the LDA value \cite{noteDC}. As a consequence, the area enclosed by the $\gamma$ sheet  decreases as well.
In Fig.~\ref{fig3} we present the same quantities shown in Fig.~\ref{fig2}, however as a function
of the temperature $T$; we find that the tetragonal SO splitting $|\Delta \delta_\lambda|$ 
increases on lowering $T$,
while $|\Delta \varepsilon_{\rm CF}|$ decreases slightly.  
In comparison with the strong dependence of $\Delta \varepsilon_{\rm CF}$ with $\Delta U$, all parameters
change weakly on lowering $T$ \cite{notetemp}.

Remarkably, these effects are to a large extent dynamical in nature \cite{Kotliar,suple}.
The zero-frequency crystal-field enhancement is given by
$\Delta \varepsilon_{\rm CF}=\Delta \Sigma^{\prime} (\infty)+\frac{1}{\pi} \int d\omega 
 \;\Delta { \Sigma^{\prime\prime}}(\omega)/{\omega}$.
The term $\Delta \Sigma^{\prime} (\infty)$ can be obtained via the static mean-field Hartree-Fock (HF) method;
in the $\Delta U=\Delta U^\prime=0$ case one can show that  $\Delta \Sigma^{\prime} (\infty) \sim \frac{1}{2}(U-5J)  p$, where  $p=n_{xy}-\frac{1}{2}(n_{xz}+n_{yz})$ is the orbital polarization.
Because of the band-width miss-match \cite{Gorelov} the LDA total polarization is  $p\sim -0.17$, i.e., negative,
despite of the positive CF splitting;  in LDA+DMFT it becomes basically zero, hence $\Delta  \Sigma^{\prime} (\infty)\sim 0$
as well. The enhancement $\Delta \varepsilon_{\rm CF}>0$ comes thus essentially from the  second term;
 it turns out, by analizing the integrand $\Delta { \Sigma^{\prime\prime}}(\omega)/{\omega}$, that it has large contributions from the lower Hubbard bands.
{The SO interaction does not affect much the CF splitting, but it slightly increases the initial
orbital polarization, from $p=-0.17$ (LDA) to $p=-0.19$ (LDA+SO)}; furthermore, it couples the $e_g$ and $b_{2g}$ orbitals, yielding a negative SO polarization $p_j\sim -0.10$, with  $p_{j}\equiv n_{3/2}-n_{1/2}$,
where $n_j$ is the average occupation of an orbital with total angular momentum $j$; 
switching on the  Coulomb interaction reduces the orbital polarization $p\sim 0$ and slightly increases $p_j\sim -0.12$. Finally, when $\Delta U>0$,  electrons are transferred from the $xy$ to the $xz$ and $yz$ bands as $\Delta \varepsilon_{\rm CF}$
decreases, yielding a negative orbital polarization $p\sim -0.11$ for $\Delta U=0.45$~eV\cite{thus}.

Returning to the FS, we find that the agreement between calculations and experiments can only be recovered if both low-symmetry Coulomb terms and correlation-enhanced SO couplings are included in the calculations. {To show this and test the robustness of our conclusion, in addition to LDA+SO+DMFT  calculations for $\Delta U=0.3$~eV  (cRPA estimate) we perform a series of model calculations. For the latter we take $\Delta \varepsilon_{\rm CF}$ in the interval [-0.08,-0.02]\,eV, 
$\Delta \lambda_{xy}$ and $\Delta \lambda_{z}$  in the intervals [0.10,0.16]\,eV and  [0.04,0.08]~eV \cite{FL}. 
These intervals estimate the possible input parameters variations and are chosen around the results in Fig.~2 for $0.3$~eV$\le\Delta U\le 0.45$~eV.
In this realistic parameter range the theoretical FS is in very good agreement with experiments \cite{Notelast}, as shown in a representative case in Fig.~\ref{fig4}.}

Our results have consequences concerning the nature of Cooper pairs.
It is often assumed that Cooper pairs can be classified as singlets or triplets
\cite{Rice95,Mackenzie04,Maeno2012}. Recently, it was pointed out that  in  Sr$_{2}$RuO$_{4}$ this scenario might break down
due to  the SO interaction \cite{Veenstra}.   
Indeed, already in LDA the SO coupling is comparable with the crystal-field splitting. 
Turning on the Coulomb interaction, for $\Delta U> 0$ we find that  {the ratio $(\lambda_{i}+\Delta\lambda_{i})/|\varepsilon_{\rm CF}+\Delta \varepsilon_{\rm CF}|$
becomes even larger than $\lambda_i/\varepsilon_{\rm CF}$.}
This points to a strong spin-orbital entanglement, which should not be neglected in studying the nature of Cooper pairs, as suggested in Refs. \cite{Veenstra,Scaffidi}. 
%%%%%%%%%%%%%%%%%%%%%%%%%%%%%%%%%%%%
\begin{figure}[t]
\begin{center}
\label{FINAL}
\includegraphics[scale=0.65]{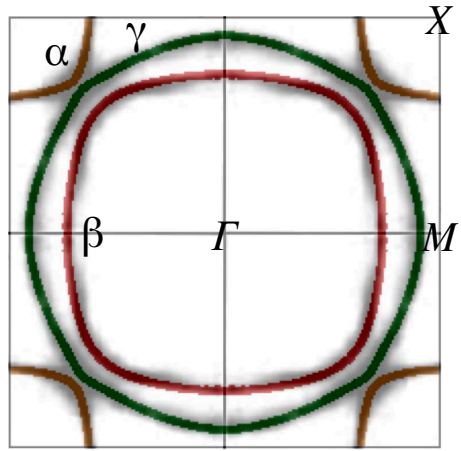}
\caption{(Color-online) 
{\label{fig4} Fermi surface ($k_z\!=\!0$) of  Sr$_{2}$RuO$_{4}$ 
  from LDA+SO+DMFT calculations  with $D_{4h}$ Coulomb terms and $(U,J)=(3.1,0.7)~$eV, $T\to 0$ limit. {Parameters: $\Delta \varepsilon_{\rm CF}\sim -0.02$~eV,  $\Delta \lambda_{xy} \sim 0.13$~eV, $\Delta  \lambda_{xz} \sim 0.08$~eV,} values approximatively corresponding to $\Delta U=3\Delta U^\prime=0.3$ eV.}
Grey density maps: experimental data from Ref.\,\onlinecite{Damascelli}.  
}
\label{default}
\end{center}
\end{figure}
%%%%%%%%%%%%%%%%%%%%%%%%%%%%%%%%%%%%

In conclusion, we investigate  in a realistic setting how  different mechanims affect the topology of the Fermi surface of Sr$_{2}$RuO$_{4}$.   
LDA calculations with spin-orbit effects describe well the
topology of the Fermi surface, but not the relative size of the Fermi sheets.
We show that adding alone the effects of the standard isotropic Coulomb interaction  
via dynamical mean-field theory does not improve (or even worsens) the agreement with experiments.
It is essential to also include the small anisotropic part of the Coulomb interaction.
Remarkably, we find that  (small) low-symmetry Coulomb  terms have a large effect at the Fermi surface.
The standard isotropic Coulomb interaction enhances the crystal-field splitting
and the spin-orbit coupling.
The Coulomb-enhanced spin-orbit coupling  shrinks the  $\beta$ sheet and extents the $\gamma$ sheet. The low-symmetry Coulomb term $\Delta U$ reduces the Coulomb crystal-field enhancement, modifying correspondingly the $\alpha$ and $\gamma$ sheets. 
To reproduce the experimental Fermi surface all these interactions are essential.
Our results support  the recent suggestions of strong spin-orbital entanglement for Cooper pairs.  
These mechanisms could be at work 
also in other multi-orbital correlated systems: other layered metallic ruthenates, iridates 
or iron-based superconductors.

We acknowledge financial support from the Deutsche Forschungsgemeinschaft through research unit FOR1346.    
The calculations were done on the J\"{u}lich Blue Gene/Q.


\begin{thebibliography}{1000}
\bibitem{DC} For $H_{\rm dc}$ we adopt an around mean-field-like form, a typical choice for correlated metals.
\bibitem{zerofreq}
We extrapolate the zero-frequency limit (shown in Fig.~\ref{fig2} and
Fig.~\ref{fig3}) from the values of
the self-energy at the first Matsubara frequency and by analytical continuation of the self-energy;
the two approaches yield similar results.
\bibitem {Spread} 
The spread $\langle|{\bm r}^2| \rangle$ is 3.79\,\AA$^{2}$ for the $xz$ and 3.55\,\AA $^{2}$ for the $xy$ orbital.
\bibitem{lastnote}
The effect of $\Delta U^\prime$, which modifies inter-orbital Coulomb terms, is weaker
than that of the intra-orbital correction $\Delta U$, hence the results of the two sets of calculations are similar. Thus only results for $\Delta U^\prime=\Delta U/3$ are shown.
\bibitem{note}
{For $(U,J)=(2.3,0.4)$~eV the changes with respect to calculations with isotropic $U$ are smaller.}
\bibitem{noteDC} 
For isotropic Coulomb interaction, $H_{dc}$ is a mere shift of the chemical potential and has no effect on the parameters. Instead, for finite $\Delta U$, $H_{dc}\propto\Delta U$. A positive (negative)  $\delta H_{dc}$
yields correspondingly a positive (negative) $\delta\Delta \varepsilon_{\rm CF}$; the other parameters are little affected by $\delta H_{\rm dc}$.
For realistic $\delta H_{dc}$ the change $\delta \Delta \varepsilon_{\rm CF}$ is, however, small; an increase of $H_{\rm dc}$ of, e.g., $9\%$, yields for $\Delta U=3\Delta U^\prime=0.45$~eV a $\Delta \varepsilon_{\rm CF}\sim -57$ meV (instead of -83~meV).  Thus the main effect of a $\delta H_{dc}\sim \pm 9\% H_{dc}$ is equivalent to that of a shift of a few tenth of meV on the right/left (depending on the sign of $\delta H_{dc}$) of the value of $\Delta U$ for which the best agreement  with experiments is reached. 
For comparison, $|\delta H_{dc}|\sim 12\%H_{dc}$ is the difference between around-mean-field limit and fully-localized limit with $n_{xy}=1,n_{xz}+n_{yz}=3$. 
Hence, our conclusions remain unaffected by $\delta H_{\rm dc}$ unless $H_{dc}$ becomes unrealistically large, yielding a large orbital polarization, to the best of our knowledge never reported experimentally.
\bibitem{notetemp} 
This, together with the Luttinger theorem, allows the $T\to 0$ extrapolation.
The strongest temperature dependence is perhaps shown by $\Delta \delta_\lambda$.
\bibitem{thus} Thus neither the SO interaction nor the crystal field splitting dominate in determining the most occupied states.
\bibitem{FL} In line with our LDA+SO+DMFT results, we assume that also for the model calculations the system is a Fermi liquid satisfying the Luttinger theorem.
\bibitem{Notelast} The agreement remains good provided that $\Delta \varepsilon_{\rm CF}$ is at least 100 meV smaller than the value obtained with isotropic $U$ and that the SO couplings are correspondingly sufficiently enhanced by Coulomb repulsion. It starts to visibly deteriorate for, e.g., $\Delta  \varepsilon_{\rm CF}\sim +0.03$~eV.
\end{thebibliography}
 \end{document}